\documentclass[12pt,draftcls,onecolumn]{IEEEtran}
\usepackage[T1]{fontenc}
\ifCLASSINFOpdf
\else
\fi
\usepackage{amsmath}
\usepackage{color}
\usepackage{amssymb}
\usepackage{graphicx}
\usepackage{subfigure}
\usepackage{cite}
\usepackage[cmintegrals]{newtxmath}
\begin{document}
\title{\huge Scalable Channel Estimation and Reflection Optimization for Reconfigurable Intelligent Surface-Enhanced OFDM Systems}
\author{Jiancheng~An,~
 Qingqing~Wu,~\emph{Member,~IEEE~}
 and~Chau~Yuen,~\emph{Fellow,~IEEE}
\thanks{J. An and C. Yuen are with the Engineering Product Development (EPD) Pillar, Singapore University of Technology and Design, Singapore 487372 (e-mail: jiancheng$\_$an@163.com; yuenchau@sutd.edu.sg).}
\thanks{Q. Wu is with the State Key Laboratory of Internet of Things for Smart City, University of Macau, Macau, 999078 (e-mail: qingqingwu@um.edu.mo).}
\vspace{-1cm}}
\maketitle
\begin{abstract}
This paper proposes a scalable channel estimation and reflection optimization framework for reconfigurable intelligent surface (RIS)-enhanced orthogonal frequency division multiplexing (OFDM) systems. Specifically, the proposed scheme firstly generates a training set of RIS reflection coefficient vectors offline. For each RIS reflection coefficient vector in the training set, the proposed scheme estimates only the end-to-end composite channel and then performs the transmit power allocation. As a result, the RIS reflection optimization is simplified by searching for the optimal reflection coefficient vector maximizing the achievable rate from the pre-designed training set. The proposed scheme is capable of flexibly adjusting the training overhead according to the given channel coherence time, which is in sharp contrast to the conventional counterparts. Moreover, we discuss the computational complexity of the proposed scheme and analyze the theoretical scaling law of the achievable rate versus the number of training slots. Finally, simulation results demonstrate that the proposed scheme is superior to existing approaches in terms of decreasing training overhead, reducing complexity as well as improving rate performance in the presence of channel estimation errors.
\end{abstract}

\IEEEpeerreviewmaketitle
\section{Introduction}
\IEEEPARstart{R}{ecently}, reconfigurable intelligent surface (RIS) and its variants have emerged as promising technologies for further improving spectrum- and energy-efficiency of wireless networks \cite{Renzo_JSAC_2020_Smart, Wu_TC_2021_Intelligent}. Specifically, RIS is a two-dimensional metasurface composed of a large number of reflecting elements, each of which is capable of rescattering incident signals by imposing an independent amplitude and/or phase shift \cite{Renzo_JSAC_2020_Smart}. In contrast to traditional active relays, RISs passively reflect impinging signals without employing any active radio-frequency (RF) chain, thus significantly reducing the power consumption as well as the transmission delay \cite{Wu_TC_2021_Intelligent}. Besides, RISs operate in a full-duplex mode without encountering severe self-interference issues \cite{Huang_JSAC_2020_Reconfigurable}. Furthermore, lightweight and low-cost features make RISs easier to install for practical deployment \cite{Wu_TWC_2019_Intelligent, An_arXiv_2021_Joint}. Due to the aforementioned merits, RIS is regarded as a competitive technology for next-generation wireless networks \cite{Zhang_JSAC_2020_Prospective}.

Orthogonal frequency division multiplexing (OFDM) is widely used in modern wireless communication systems for its robustness against frequency-selective fading channels. Recently, the authors of \cite{Yang_TC_2020_Intelligent, Zheng_WCL_2020_Intelligent} studied the channel estimation and reflection optimization for RIS-enhanced OFDM systems. However, on one hand, channel state information (CSI) acquisition for the growing number of reflected links therein requires plenty of pilots proportional to the number of RIS elements, which poses a substantial challenge due to RIS's passive structure \cite{Jensen_ICASSP_2020_AN, Wei_TC_2021_Channel}. Although recent so-called improved methods have been proposed to reduce pilot overheard by exploiting the correlation and sparsity of reflected channels, the number of pilots is still proportional to that of RIS elements \cite{Wang_TWC_2020_Channel, Wang_SPL_2020_Compressed}. On the other hand, RIS reflection needs to be optimized jointly with the active components coexisting in networks, resulting in non-trivial optimization problems, which can be solved by invoking alternating optimization (AO) algorithm and successive convex approximation technique of high complexity \cite{Di_TVT_2020_Practical, Wu_TC_2020_Beamforming, Han_TVT_2019_Large, An_WCL_2021_The}. In a nutshell, the rigid limitation on the training overhead and forbidden complexity renders it difficult for existing solutions to be applied to rapidly time-varying channels and large-scale RIS deployment.

To address the above issue, this paper proposes a scalable channel estimation and reflection optimization framework for maximizing the achievable rate of RIS-enhanced OFDM systems. In contrast to existing solutions for RIS-enhanced OFDM systems \cite{Yang_TC_2020_Intelligent, Zheng_WCL_2020_Intelligent, An_arXiv_2021_Reconfigurable}, the proposed scheme firstly generates a training set of RIS reflection coefficient vectors offline. For each given RIS reflection coefficient vector in the training set, the composite channel estimation and transmit power allocation can be performed in the same way as the conventional OFDM systems without RIS. As a consequence, the RIS reflection optimization is simplified by searching for the optimal one maximizing the achievable rate from the pre-designed training set. Moreover, we discuss the computational complexity and analyze the theoretical scaling law of the achievable rate versus the number of training slots. Finally, numerical simulations evaluate the rate performance of the proposed scheme and verify our theoretical analysis. Simulation results demonstrate that our proposed scheme is capable of striking flexible trade-offs between the training overhead and the received power thus maximizing the effective achievable rate.
\section{System Model}\label{s2}

\begin{figure}[!t]
	\centering
	\includegraphics[width=6cm]{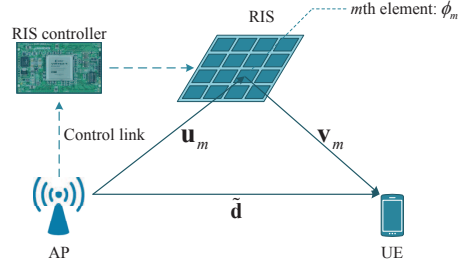}
	\caption{Schematic of a RIS-enhanced OFDM system.}
	\label{fig1}
	\vspace{-0.6cm}
\end{figure}

As illustrated in Fig. \ref{fig1}, we consider a RIS-enhanced OFDM communication system, where a RIS equipped with $M$ reflecting elements is deployed to enhance the communication link between an access point (AP) and an user equipment (UE), both of which are equipped with a single antenna. The RIS is connected to a smart RIS controller, which is capable of manipulating RIS's electromagnetic response dynamically for achieving desired signal reflection. Let ${\boldsymbol{\phi}} = {\left[ {{\phi _1}, \cdots ,{\phi _M}} \right]^T} \in {{\mathbb{C}}^{M \times 1}}$ denote the reflection coefficient vector at the RIS, where ${\phi _m}$ denotes the reflection coefficient of the $m$th RIS element. In order to maximize the received power and simplify hardware design, we assume $\left| {{\phi _m}} \right| = 1$ in this paper \cite{Wu_TC_2020_Beamforming}. Moreover, we consider the quasi-static frequency-selective fading channels and adopt the time-division duplex mode. Specifically, let ${\bf{\tilde d}} \in {{\mathbb{C}}^{{L_d} \times 1}}$ denote the ${L_{d}}$-tap baseband equivalent channel from the AP to the UE. Similarly, let ${{\bf{u}}_m} \in {{\mathbb{C}}^{{L_{u}} \times 1}}$ and ${{\bf{v}}_m} \in {{\mathbb{C}}^{{L_{v}} \times 1}}$ represent the baseband equivalent channels of the ${L_{u}}$-tap AP-RIS link and the ${L_{v}}$-tap RIS-UE link, respectively, associated with the $m$th reflecting element. Furthermore, let ${\bf{d}} = {\left[ {{{{\bf{\tilde d}}}^T},{{\bf{0}}_{1 \times \left( {N - {L_d}} \right)}}} \right]^T} \in {{\mathbb{C}}^{N \times 1}}$ and ${{\bf{r}}_m} = {\left[ {{{\left( {{{\bf{u}}_m} * {{\bf{v}}_m}} \right)}^T},{{\bf{0}}_{1 \times \left( {N - {L_{r}}} \right)}}} \right]^T} \in {{\mathbb{C}}^{N \times 1}}$ denote the zero-padded direct channel and cascaded reflected channel via the $m$th element, respectively, where we have ${L_r} = {L_u} + {L_v} - 1$, generally satisfying ${L_r} \ge {L_d}$. Hence, the composite channel spanning from the AP to the UE is given by ${\bf{h}} = {\bf{d}} + \sum\nolimits_{m = 1}^M {{\phi _m}} {{\bf{r}}_m} = {\bf{d}} + {\bf{R}}{\boldsymbol{\phi}}$, where we have ${\bf{R}} = \left[ {{{\bf{r}}_1}, \cdots ,{{\bf{r}}_M}} \right]$.

Furthermore, the total bandwidth assigned to the UE is equally divided into $N$ orthogonal subcarriers. Let ${\bf{p}} = {\left[ {{p_0}, \cdots ,{p_{N - 1}}} \right]^T} \in {{\mathbb{R}}^{N \times 1}}$, where ${p_n} \ge 0$ denotes the power allocated to the $n$th subcarrier at the AP. Assume that the aggregate transmit power available at the AP is ${P_{DL}}$. Thus, the power allocation solution should satisfy $\sum\nolimits_{n = 0}^{N - 1} {{p_n}} \le {P_{DL}}$. At the transmitter, the OFDM symbol is first transformed into the time domain via an $N$-point inverse discrete Fourier transform, and then appended by a cyclic prefix (CP) of length ${N_{CP}}$, which is assumed to be longer than the maximum delay spread of all reflected channels, i.e., ${N_{CP}} \ge {L_{r}}$. As a result, the achievable rate of RIS-enhanced OFDM systems in terms of bits per second per Hertz is given by \cite{Yang_TC_2020_Intelligent}\begin{small}\begin{align}
R\left( {{\bf{p}},{\boldsymbol{\phi}} } \right) = \frac{1}{{N + {N_{CP}}}}\sum\limits_{n = 0}^{N - 1} {{{\log }_2}\left( {1 + \frac{{{{\left| {{\bf{f}}_n^H{\bf{d}} + {\bf{f}}_n^H{\bf{R}}{\boldsymbol{\phi}} } \right|}^2}{p_n}}}{{\sigma _w^2}}} \right)},
\end{align}\end{small}where ${\bf{f}}_n^H$ denotes the $n$th row of the discrete Fourier transform (DFT) matrix ${\bf{F}}$, $\sigma _w^2$ denotes the average noise power at the UE.

In this paper, we aim to maximize the achievable rate of RIS-enhanced OFDM systems by jointly optimizing the transmit power allocation at the AP and the reflection pattern at the RIS. Accordingly, the optimization problem is formulated by\begin{small}\begin{align}\label{eq3}
\begin{array}{*{20}{l}}
{\mathop {\max }\limits_{{\bf{p}},{\boldsymbol{\phi}} } }&{\sum\nolimits_{n = 0}^{N - 1} {{{\log }_2}\left( {1 + {{{{\left| {{\bf{f}}_n^H{\bf{d}} + {\bf{f}}_n^H{\bf{R}}{\boldsymbol{\phi}} } \right|}^2}{p_n}} \mathord{\left/
 {\vphantom {{{{\left| {{\bf{f}}_n^H{\bf{d}} + {\bf{f}}_n^H{\bf{R}}{\boldsymbol{\phi}} } \right|}^2}{p_n}} {\sigma _w^2}}} \right.
 \kern-\nulldelimiterspace} {\sigma _w^2}}} \right)} }\\
{s.t.}&{\sum\nolimits_{n = 0}^{N - 1} {{p_n}} \le {P_{DL}},\ {p_n} \ge 0,\ n = 0, \cdots ,N - 1;}\\
{}&{\left| {{\phi _m}} \right| = 1,\ m = 1, \cdots ,M.}
\end{array}
\end{align}\end{small}Note that in order to design ${\bf{p }}$ and ${\boldsymbol{\phi}}$, the AP needs the prior knowledge of CSI, i.e., $\left\{ {{\bf{d}},{{\bf{r}}_1}, \cdots ,{{\bf{r}}_m}} \right\}$. Whereas the pilot overhead of existing channel estimation approaches is proportional to the number of RIS elements, which renders them inapplicable for rapidly time-varying channels in practice \cite{Zheng_WCL_2020_Intelligent, Jensen_ICASSP_2020_AN}. Besides, the joint optimization of the transmit power allocation and reflection coefficient vector suffers from high computational complexity \cite{Yang_TC_2020_Intelligent}. Therefore, it is advisable to design a low-complexity overhead-aware scheme to strike flexible trade-offs between pilot overhead, computational complexity, and achievable rate according to the specific quality-of-service requirements.

\section{The Proposed Scalable Framework}\label{s3}
\begin{figure}[!t]
	\centering
	\includegraphics[width=7cm]{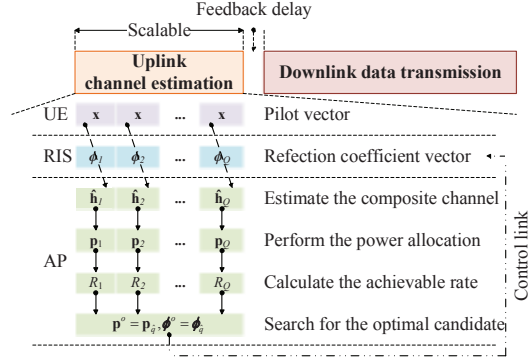}
	\caption{Illustration of the proposed scalable transmission protocol.}
	\label{fig2}
	\vspace{-0.4cm}
\end{figure}

In order to tackle the aforementioned issues, we propose in this section a channel estimation and reflection optimization framework for RIS-enhanced OFDM systems. The transmission protocol of the proposed solution is portrayed in Fig. \ref{fig2}, while the detailed procedures are elaborated as follows.
\subsection{Estimation of the Composite Channel}\label{s3-1-1}
In contrast to existing approaches, which estimate the direct and reflected channels separately, the proposed scheme first generates a training set containing $Q$ RIS reflection coefficient vectors from the legitimate solution set, denoted by $\left\{ {{{\boldsymbol{\phi}} _1}, \cdots ,{{\boldsymbol{\phi}} _Q}} \right\}$. The specific method for generating the training set will be discussed later in Section \ref{sec4-2}. For each given RIS reflection pattern in the training set, only the composite channel needs to be estimated. Specifically, let ${\bf{x}} \in {{\mathbb{C}}^{N \times 1}}$ satisfying ${\left\| {\bf{x}} \right\|^2} = 1$ denote the normalized pilot vector transmitted from the UE. Thus, at the $q$th training slot, the baseband signal ${{\bf{y}}_q} \in {{\mathbb{C}}^{N \times 1}}$ received by the AP in the frequency domain is given by\begin{small}\begin{align}
{{\bf{y}}_q} = \left[ {{\bf{F}}\left( {{\bf{d}} + {\bf{R}}{{\boldsymbol{\phi}} _q}} \right)} \right] \circ \sqrt {{P_{UL}}} {\bf{x}} + {{\bf{z}}_q} = \sqrt {{P_{UL}}} {\bf{XF}}{{\bf{h}}_q} + {{\bf{z}}_q},
\end{align}\end{small}where we have ${\bf{X}} = {\text{diag}}\left( {\bf{x}} \right)$, $ \circ $ represents the Hadamard product, while ${{\bf{h}}_q} = {\bf{d}} + {\bf{R}}{{\boldsymbol{\phi}} _q}$ denotes the composite channel combining $1$ direct and $M$ reflected channels at the $q$th training slot, ${{\bf{z}}_q} \sim {\mathcal{CN}}\left( {{\bf{0}},\sigma _z^2{{\bf{I}}_N}} \right)$ denotes the noise vector at the $q$th training slot.
 
Therefore, the least-square estimate of the composite channel at the $q$th training slot can be readily obtained by\begin{small}\begin{align}\label{eq5}
{{\bf{\tilde h}}_q} = \frac{1}{{N\sqrt {{P_{UL}}} }}{{\bf{F}}^H}{{\bf{X}}^{ - 1}}{{\bf{y}}_q}.
\end{align}\end{small}With the prior information of the channel order, the composite channel estimate can be further improved by yielding ${{\bf{\hat h}}_q} = {\left[ {{{{\bf{\tilde h}}}_{q,1}}, \cdots ,{{{\bf{\tilde h}}}_{q,{L_r}}},{0_{1 \times \left( {N - {L_r}} \right)}}} \right]^T}$.
\subsection{Transmit Power Allocation at the AP}\label{s3-1-2}
After obtaining the composite channel estimate ${{\bf{\hat h}}_q}$ at the $q$th training slot, the original optimization problem in (\ref{eq3}) degrades to\begin{small}\begin{align}
\begin{array}{*{20}{l}}
{\mathop {\max }\limits_{{{\bf{p}}_q}} }&{\sum\nolimits_{n = 0}^{N - 1} {{{\log }_2}\left( {1 + {{{{\left| {{\bf{f}}_n^H{{{\bf{\hat h}}}_q}} \right|}^2}{p_{q,n}}} \mathord{\left/
 {\vphantom {{{{\left| {{\bf{f}}_n^H{{{\bf{\hat h}}}_q}} \right|}^2}{p_{q,n}}} {\sigma _w^2}}} \right.
 \kern-\nulldelimiterspace} {\sigma _w^2}}} \right)} }\\
{s.t.}&{\sum\nolimits_{n = 0}^{N - 1} {{p_{q,n}}} \le {P_{DL}},\ {p_{q,n}} \ge 0,\ n = 0, \cdots ,N - 1.}
\end{array}
\end{align}\end{small}which turns out to be the power allocation problem in conventional OFDM systems operating without RIS. It is known that the optimal transmit power allocation ${\bf{p}}_q$ follows the classic water-filling structure, which can be expressed as \cite{Yang_TC_2020_Intelligent}\begin{small}\begin{align}\label{eq8}
{p_{q,n}} = \max \left( {c - {{\sigma _w^2} \mathord{\left/
 {\vphantom {{\sigma _w^2} {{{\left| {{\bf{f}}_n^H{{{\bf{\hat h}}}_q}} \right|}^2}}}} \right.
 \kern-\nulldelimiterspace} {{{\left| {{\bf{f}}_n^H{{{\bf{\hat h}}}_q}} \right|}^2}}},0} \right),
\end{align}\end{small}where $c$ denotes the cut-off power threshold that enables $\sum\nolimits_{n = 0}^{N - 1} {{p_{q,n}}} = {P_{DL}}$.

Therefore, the expected achievable rate with given RIS reflection coefficient vector ${{\boldsymbol{\phi}} _q}$ is given by\begin{small}\begin{align}
{R_q} = \frac{1}{{N + {N_{CP}}}}\sum\limits_{n = 0}^{N - 1} {{{\log }_2}\left( {1 + \frac{{{{\left| {{\bf{f}}_n^H{{{\bf{\hat h}}}_q}} \right|}^2}{p_{q,n}}}}{{\sigma _w^2}}} \right)}.
\end{align}\end{small}
\vspace{-0.6cm}
\subsection{Reflection Optimization at the RIS}\label{s3-1-3}
For different training slot, we configure the RIS based on the reflection coefficient vector ${{\boldsymbol{\phi}} _q}$ and repeat performing the composite channel estimation and the transmit power allocation. Following this, the reflection optimization is simplified to searching for the optimal one that maximizes the achievable rate from $Q$ candidates, which can be expressed by\begin{small}\begin{align}
\hat q = \arg \mathop {\max }\limits_{q = 1, \cdots ,Q} {R_q}.
\end{align}\end{small}Once that the optimal index $\hat q$ is obtained, accordingly the resultant RIS reflection coefficient vector and transmit power allocation solution are also determined by yielding $ {{\boldsymbol{\phi}}^o} = {{\boldsymbol{\phi}} _{\hat q}}$ and ${{\bf{p}}^o} = {\bf{p}}_{\hat q}$, respectively. Fig. \ref{fig2} also summarizes the general steps of the proposed scheme.

Note that the proposed framework overcomes the poor scalability of the existing channel estimation and reflection optimization solutions. Specifically, the pilot overhead of the proposed framework is independent of the number of RIS reflecting elements and thus can be dynamically adjusted for striking flexible trade-offs between the training overhead and system performance.
\section{Performance Analysis}
In this section, we will discuss the computational complexity and analyze the achievable rate of the above proposed scheme.
\subsection{Computational Complexity Analysis}
Let us first consider the channel estimation for RIS-enhanced OFDM systems. The conventional methods need to estimate one direct channel and $M$ reflected channels associated with $N$ subcarriers, where the computational complexity required is $2N\left( {M + 1} \right)$, in terms of the number of real-valued multiplications \cite{Yang_TC_2020_Intelligent, An_WCL_2021_The}. By contrast, the proposed solution only estimates one composite channel at each training slot, resulting in a complexity of $2NQ$. As for the reflection optimization, the computational complexity of the joint transmit power allocation and reflection optimization is ${N_i}M\left( {4M{L_r} + N\left( {6{L_r} + 2} \right) + 4} \right)$, where ${N_{i}}$ represents the number of iterations of the AO algorithm, while ${4M{L_r}}$, ${N\left( {6{L_r} + 1} \right)}$, and $4$ denote the complexity for calculating ${\bf{R}} {\boldsymbol{\phi}} $, obtaining ${\bf{p}}$, and optimizing one element of ${\boldsymbol{\phi}}$ \cite{Yang_TC_2020_Intelligent, Wu_TC_2020_Beamforming}, respectively, at each iteration. Note that here we adopt the method of optimizing a single reflection phase shift by fixing the remaining $M-1$ phase shifts in each iteration because the complexity order of the successive convex approximation (SCA) technique, i.e., ${\mathcal{O}}\left( {{M^{4.5}}{N^{3.5}}} \right)$, is extremely high \cite{Yang_TC_2020_Intelligent}. By contrast, the proposed solution only carries out the transmit power allocation and then calculates the achievable rate at each training slot. Hence, the complexity of the proposed framework is $QN\left( {6{L_r} + 2} \right)$, which is significantly reduced compared to existing counterparts.

\subsection{Achievable Rate Analysis}\label{sec4-2}
Next, we will evaluate the performance of the proposed framework by analyzing the theoretical scaling law of the achievable rate of RIS-enhanced OFDM systems versus the number of training slots. In this paper, we adopt the straightforward random configuration for generating the training set of $Q$ RIS reflection coefficient vectors. More sophisticated training set generation methods and their performance analysis will be reserved for our future research. Specifically, each RIS phase shift at each training slot is uniformly and randomly distributed in $\left[ {0,2\pi } \right)$, i.e., $\angle {\phi _{q,n}} \sim {\mathcal{U}}\left[ {0,2\pi } \right)$. We assume that the CSI obtained at each training slot is error-free. As a result, the ergodic achievable rate adopting the proposed framework is given by\begin{small}\begin{align}\label{eq9}
{\mathbb{E}}\left( R \right) =& {\mathbb{E}}\left\{ {\mathop {\max }\limits_{q = 1, \cdots ,Q} \left[ {\frac{1}{{N + {N_{CP}}}}\sum\limits_{n = 0}^{N - 1} {{{\log }_2}\left( {1 + \frac{{{{\left| {{\bf{f}}_n^H{{\bf{h}}_q}} \right|}^2}{p_{q,n}}}}{{\sigma _w^2}}} \right)} } \right]} \right\} \notag \\
 \overset{M \to \infty}{\simeq}& {\mathbb{E}}\left\{ {\mathop {\max }\limits_{q = 1, \cdots ,Q} \left[ {\frac{N}{{N + {N_{CP}}}}{{\log }_2}\left( {1 + \frac{{{P_{DL}}{{\left\| {{{\bf{h}}_q}} \right\|}^2}}}{{N\sigma _w^2}}} \right)} \right]} \right\} \notag \\
 \le& \frac{N}{{N + {N_{CP}}}}{\log _2}\left( {1 + \frac{{{P_{DL}}}}{{N\sigma _w^2}}{\mathbb{E}}\left[ {\mathop {\max }\limits_{q = 1, \cdots ,Q} {{\left\| {{{\bf{h}}_q}} \right\|}^2}} \right]} \right).
\end{align}\end{small}Furthermore, let ${\kappa _d}$ and ${\kappa _r} = {\kappa _u}{\kappa _v}$ denote the proportions of the line-of-sight (LoS) component of the direct link and reflected links, respectively, satisfying $0 \le {\kappa _d},{\kappa _r} \le 1$. Let $\rho _d^2$ and $\rho _r^2 = \rho _u^2\rho _v^2$ denote the average power of the direct link and reflected links, respectively. Since it is difficult to obtain the closed-form expression of (\ref{eq9}), \emph{Proposition 1} provides an asymptotic expression of the average channel gain in (\ref{eq9}) as ${\kappa _r} \to 1$, which can be readily satisfied when the RIS is deployed at a position enabling the LoS propagation with the AP/UE.

\emph{Proposition 1:} As $M \to \infty $ and ${\kappa _r} \to 1$, it holds that\begin{small}\begin{align}
\mathbb{E}\left[ {\mathop {\max }\limits_{q = 1, \cdots ,Q} {{\left\| {{{\bf{h}}_q}} \right\|}^2}} \right] \to& \left( {{\kappa _d}\rho _d^2 + M{\kappa _r}\rho _r^2} \right)\left( {\log Q + C} \right) \notag\\
 &+ \left( {1 - {\kappa _d}} \right)\rho _d^2 + M\left( {1 - {\kappa _r}} \right)\rho _r^2,
\end{align}\end{small}where $C \approx 0.57722 \ldots $ is the Euler-Mascheroni constant.

\emph{Proof:} As ${\kappa _r} \to 1$, the average channel gain in (\ref{eq9}) can be approximated by\begin{small}\begin{align}\label{eq11}
{\mathbb{E}}\left( {\mathop {\max }\limits_{q = 1, \cdots ,Q} {{\left\| {{{\bf{h}}_q}} \right\|}^2}} \right) \simeq {\mathbb{E}}\left( {\mathop {\max }\limits_{q = 1, \cdots ,Q} {{\left| {{h_{q,1}}} \right|}^2}} \right) + \sum\limits_{n = 2}^{{L_r}} {{\mathbb{E}}\left( {{{\left| {{h_{1,n}}} \right|}^2}} \right)}.
\end{align}\end{small}Based on the Lindeberg-Levy central limit theorem \cite{Wu_TC_2020_Beamforming}, as $M \to \infty $, we have\begin{small}\begin{align}
{h_{q,1}}\sim&{\cal C}{\cal N}\left( {0,{\kappa _d}\rho _d^2 + M{\kappa _r}\rho _r^2} \right),\\
{h_{1,n}}\sim&{\cal C}{\cal N}\left( {0,\frac{{1 - {\kappa _d}}}{{{L_r} - 1}}\rho _d^2 + M\frac{{1 - {\kappa _r}}}{{{L_r} - 1}}\rho _r^2} \right),
\end{align}\end{small}where the normalized channel vector and the uniform power delay profile are considered.

Furthermore, the distribution of the $Q$th order statistics $\mathop {\max }\limits_{q = 1, \cdots ,Q} \left( {{{\left| {{h_{q,1}}} \right|}^2}} \right)$ in (\ref{eq11}) is equivalent to \cite{An_WCL_2021_The}\begin{small}\begin{align}
\mathop {\max }\limits_{q = 1, \cdots ,Q} \left( {{{\left| {{h_{q,1}}} \right|}^2}} \right) = \left( {\kappa _d^2\rho _d^2 + M\kappa _r^2\rho _r^2} \right)\sum\nolimits_{j = 1}^Q {\frac{{\gamma _j}}{{Q - j + 1}}},
\end{align}\end{small}where ${\gamma _j}$'s, $j = 1, \cdots ,Q$ are i.i.d. standard exponential random variables. Therefore, the expectation of the $Q$th order statistics $\mathop {\max }\limits_{q = 1, \cdots ,Q} \left( {{{\left| {{h_{q,1}}} \right|}^2}} \right)$ in (\ref{eq11}) can be obtained by\begin{small}\begin{align}\label{eq12}
{\mathbb{E}}\left( {\mathop {\max }\limits_{q = 1, \cdots ,Q} {{\left| {{h_{q,1}}} \right|}^2}} \right) =& {\mathbb{E}}\left[ {\left( {\kappa _d^2\rho _d^2 + M\kappa _r^2\rho _r^2} \right)\sum\nolimits_{j = 1}^Q {\frac{{{\gamma _j}}}{{Q - j + 1}}} } \right] \notag\\
 =& \left( {\kappa _d^2\rho _d^2 + M\kappa _r^2\rho _r^2} \right)\sum\nolimits_{j = 1}^Q {\frac{1}{{Q - j + 1}}} \notag\\
 \overset{Q \to \infty}{\simeq}& \left( {\kappa _d^2\rho _d^2 + M\kappa _r^2\rho _r^2} \right)\left( {\log Q + C} \right).
\end{align}\end{small}Upon substituting (\ref{eq12}) into (\ref{eq11}), the proof is completed. $\hfill\blacksquare$

\emph{Proposition 1} theoretically demonstrates that the proposed framework is capable of adapting the number of training slots for striking flexible trade-offs between the pilot overhead and rate performance, which is essentially different from existing solutions with fixed pilot overhead.
\section{Simulation Results}\label{s5}
In this section, numerical results are provided to evaluate the achievable rate performance of the proposed framework. We assume that the RIS with a uniform rectangular array is deployed on the $x$-$z$ plane to enhance the OFDM communications. The element spacing is $\lambda /8$ and the number of elements along the $x$-axis is fixed to $10$. The height of the AP and the RIS are set to $10$ m, while the horizontal length of the AP-RIS link is set to $50$ m. We assume that the RIS is deployed right above the UE. Moreover, the multi-path Rician fading model is adopted for all channels involved in Fig. \ref{fig1} \cite{Yang_TC_2020_Intelligent}. Specifically, the maximum delay spread of the AP-UE, AP-RIS and RIS-UE links are set to ${L_d} = 3$, ${L_u} = 1$, ${L_v} = 5$, respectively. The Rician factors of the AP-UE, AP-RIS and RIS-UE links are set to ${\gamma _d} = 0$ dB, ${\gamma _u} = 5$ dB, ${\gamma _v} = 3$ dB, respectively. Accordingly the corresponding LoS proportions can be obtained by yielding ${\kappa _{d/r}} = {{{\gamma _{d/r}}} \mathord{\left/
 {\vphantom {{{\gamma _{d/r}}} {\left( {{\gamma _{d/r}} + {L_{d/r}} - 1} \right)}}} \right.
 \kern-\nulldelimiterspace} {\left( {{\gamma _{d/r}} + {L_{d/r}} - 1} \right)}}$. Furthermore, the distance-dependent path loss of each link is modeled by ${\rho ^2} = {C_0}{d^{ - \alpha }}$, where ${C_0} = - 30$ dB denotes the path loss at the reference distance of $1$ m, while $d$ and $\alpha$ denote the transmission distance and the path loss exponent, respectively, of the corresponding link. In our simulations, the path loss exponents of the AP-UE, AP-RIS and RIS-UE links are set to be ${\alpha _d} = 3.5$, ${\alpha _u} = 2.2$, and ${\alpha _v} = 2.8$, respectively \cite{An_arXiv_2021_Joint}. In addition, the number of OFDM subcarriers is set to $N = 128$, while the CP length is set to ${N_{CP}} = 8$. The downlink transmit power at the AP is set to ${P_{DL}} = 10$ dBm, while the values of average noise power at the AP and UE are set to $\sigma _z^2 = - 100$ dBm and $\sigma _w^2 = - 90$ dBm, respectively.

\begin{figure}[!t]
	\centering
	\subfigure[\label{fig31}]{\includegraphics[width=4.37cm]{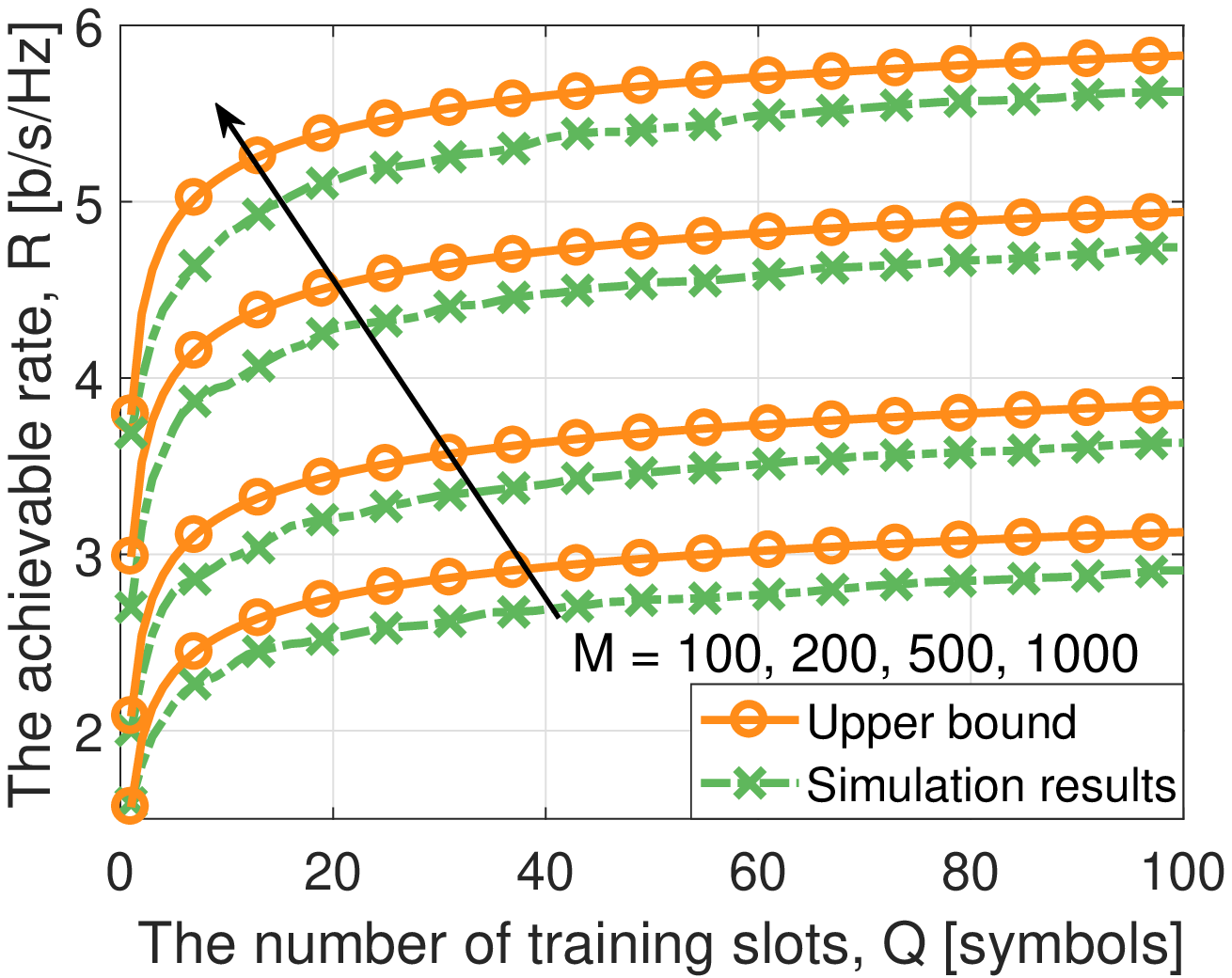}}
	\subfigure[\label{fig32}]{\includegraphics[width=4.37cm]{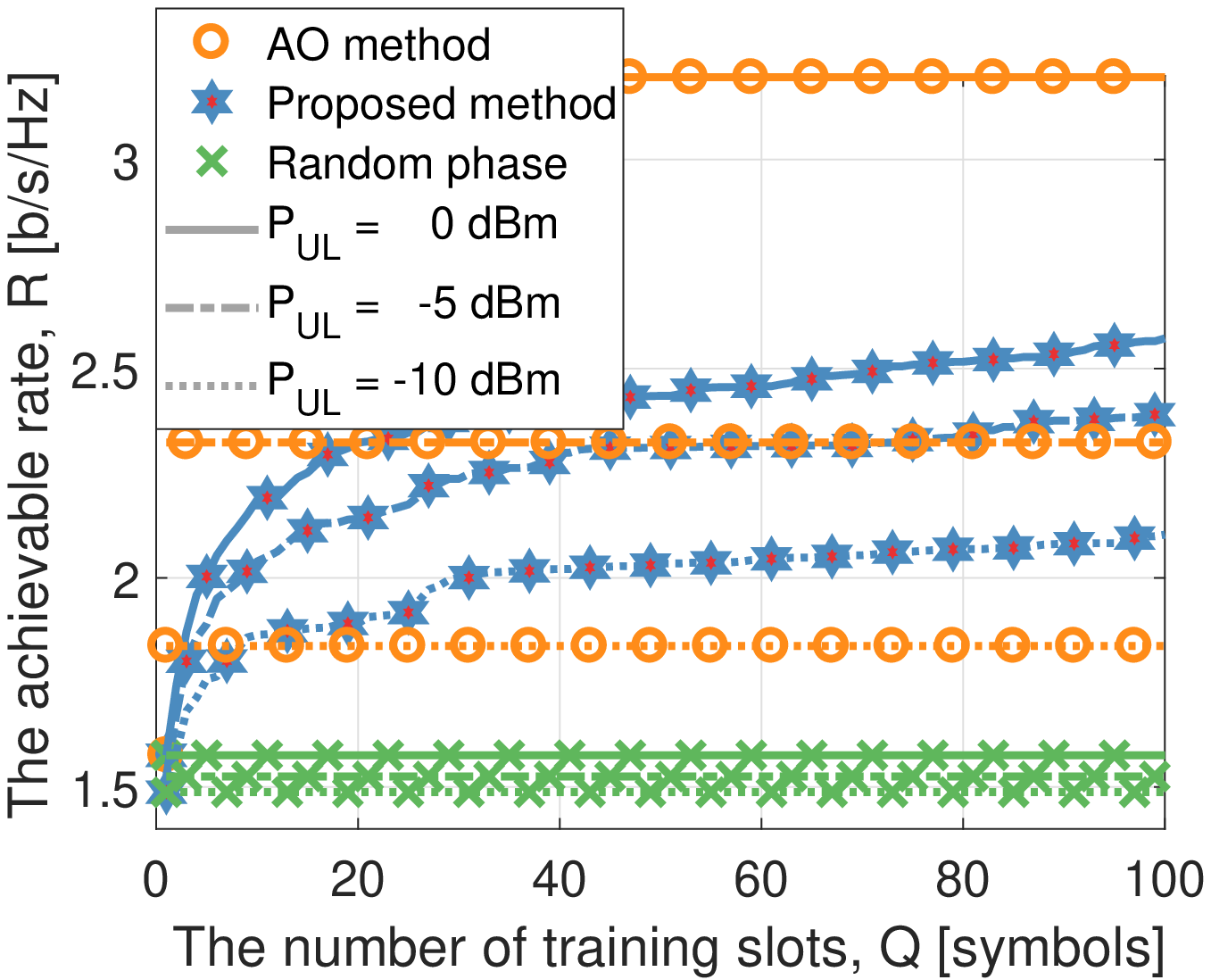}}
	\caption{\color{black}{(a) The achievable rate versus the number of training slots (perfect CSI); (b) The achievable rate versus the number of training slots (imperfect CSI, $M = 100$).}}
	\vspace{-0.5cm}
\end{figure}

First, Fig. \ref{fig31} evaluates the achievable rate performance of the proposed scheme without considering any channel estimation errors, where four RIS sizes, $M = 100,200,500,1000$, are considered. It is evident to observe that the achievable rate adopting the proposed scheme increases with the number of reflecting elements. Additionally, Fig. \ref{fig31} demonstrates that the achievable rate of the proposed scheme increases logarithmically with the number of training slots, thus striking flexible trade-offs between training overhead and achievable rate performance. Furthermore, it can be seen from Fig. \ref{fig31} that \emph{Proposition 1} offers a tight upper bound of the achievable rate versus the number of training slots. Specifically, there is only a rate gap of less than $0.2$ b/s/Hz for all the considered RIS's sizes.

Next, Fig. \ref{fig32} compares the rate performance of the proposed scheme to other counterparts under imperfect CSI, where we have $M = 100$. In our simulation, we consider two benchmark schemes: \emph{1)} the AO method characterizing the near-optimal performance \cite{Yang_TC_2020_Intelligent}, and \emph{2)} the random phase shift. In addition, three values of average pilot power are adopted for uplink training, i.e., $ {P_{UL}} = 0$ dBm, $- 5$ dBm, $- 10$ dBm. The DFT-based reflection coefficient configuration is invoked for improving the channel estimation performance \cite{Jensen_ICASSP_2020_AN, Zheng_WCL_2020_Intelligent}. As can be seen from Fig. \ref{fig32}, the proposed scheme degenerates into the random phase shift method when $Q = 1$. Furthermore, compared to the AO method, the proposed scheme has a moderate performance penalty under high pilot power. For example, about $0.6$ b/s/Hz rate erosion is observed when we adopt a fair training symbol of $Q = 100$ and ${P_{UL}} = 0$ dBm. Nevertheless, for a moderate and low pilot power, the proposed scheme might outperform the AO algorithm. For example, given ${P_{UL}} = - 10$ dBm, the proposed scheme outperforms the AO algorithm for $Q \ge 15$. Note that the AO method still requires $\left( {M + 1} \right)$ pilots to obtain such inaccurate CSI. Specifically, the estimation errors of the direct channel and all reflected channels will severely deteriorate the performance of the AO algorithm. By contrast, the estimation error of the composite channel has less impact on the performance of the proposed scheme, owing to the fact that the proposed scheme at least operates near its optimal performance even with selection bias. Therefore, the proposed scheme is more robust to channel estimation errors, as implicitly shown in Fig. \ref{fig32}.

\begin{figure}[!t]
	\centering
	\subfigure[\label{fig41}]{\includegraphics[width=4.37cm]{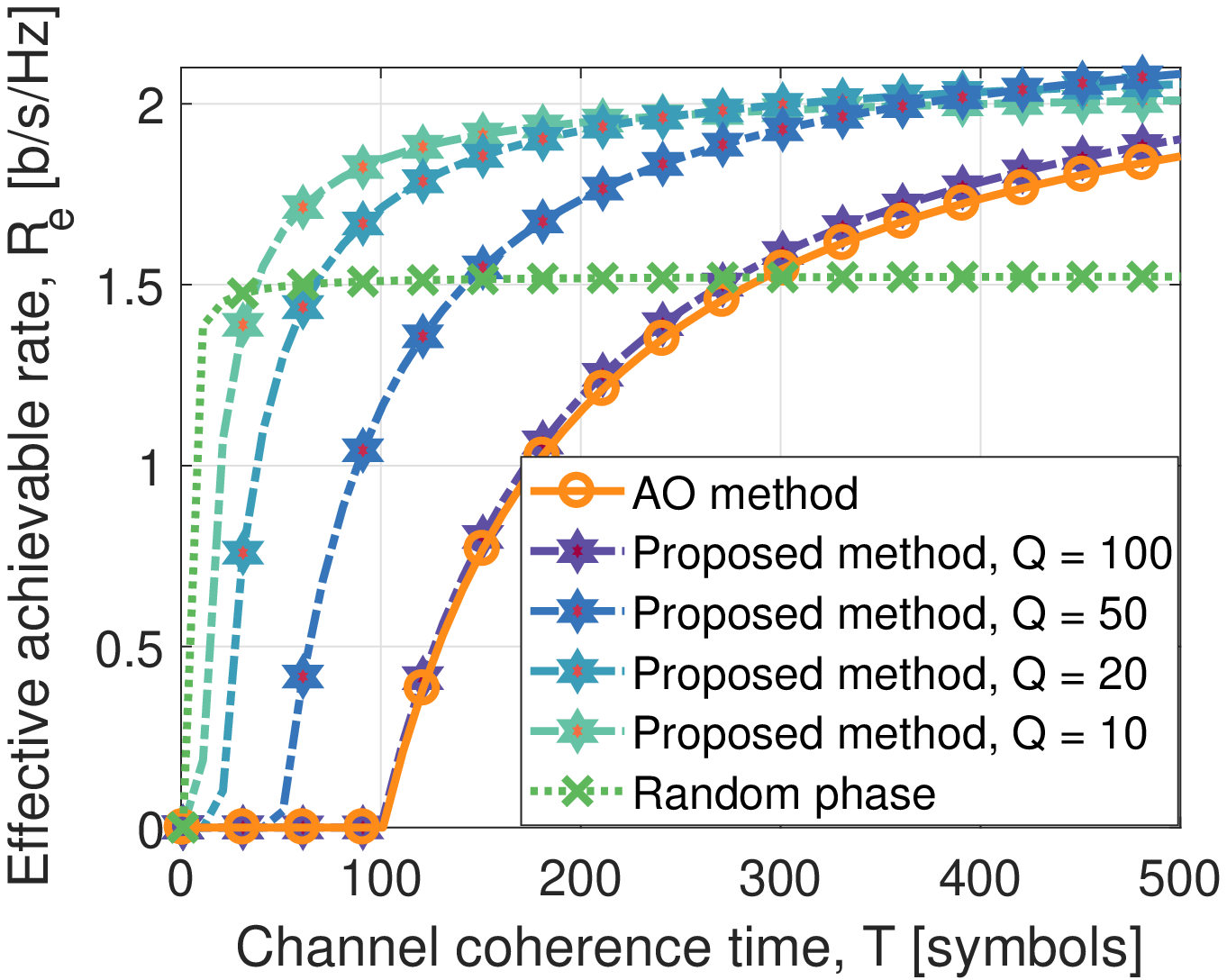}}
	\subfigure[\label{fig42}]{\includegraphics[width=4.37cm]{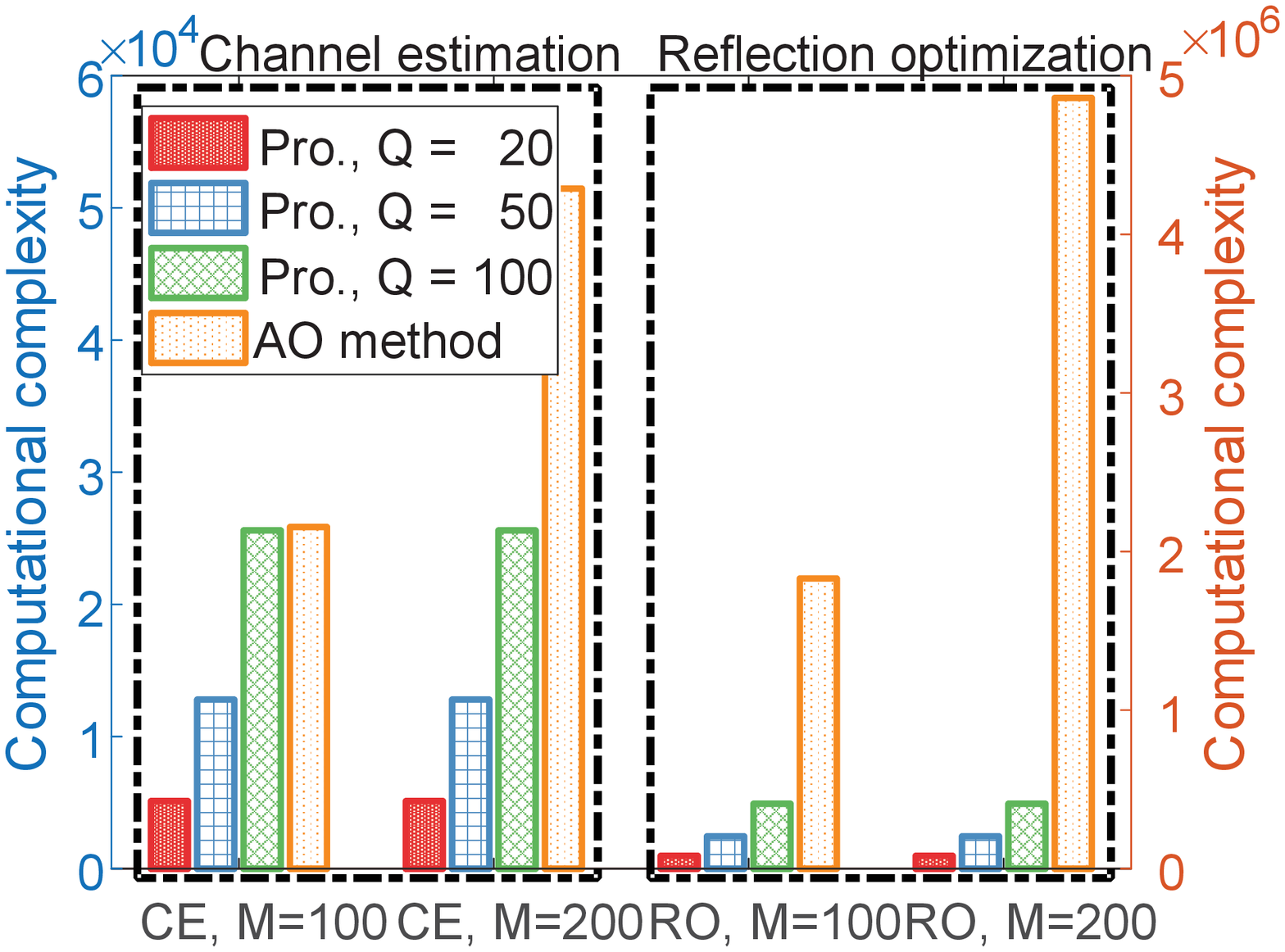}}
	\caption{\color{black}{(a) The effective achievable rate versus the channel coherence time (imperfect CSI, $M = 100$, ${P_{UL}} = - 5$ dBm); (b) The complexity comparison of the proposed scheme to the AO method (${N_{i}} = 3$).}}
	\vspace{-0.6cm}
\end{figure}

Furthermore, Fig. \ref{fig41} shows the effective achievable rate of different schemes with $M = 100$ and ${P_{UL}} = - 5$ dBm. Specifically, the effective achievable rate is defined by ${R_e} = \left( {1 - {\tau \mathord{\left/
 {\vphantom {\tau T}} \right.
 \kern-\nulldelimiterspace} T}} \right)R$, where $\tau$ and $T$ represent the training overhead and channel coherent time, respectively, in terms of symbols. We have $\tau = 1$, $\tau = Q$, and $\tau = M + 1$ for the random phase shift method, the proposed scheme and the AO method, respectively. As can be seen from Fig. \ref{fig41}, when considering the rapidly time-varying channels having short channel coherence time, the AO algorithm behaves quite poorly in terms of the effective achievable rate, because it costs excessive time to probe all reflected channels and thus limits the time for data transmission. By contrast, the proposed scheme can flexibly adjust the number of training slots $Q$, thus maximizing the effective achievable rate under different values of channel coherent time. For example, it is suggested to choose $Q = 10$ for $50 \le T < 200$ and $Q = 20$ for $200 \le T < 350$, respectively, so as to optimize the effective achievable rate in a global sense. Therefore, the proposed scalable framework constitutes a competitive transmission protocol for rapidly time-varying communications.

Finally, Fig. \ref{fig42} compares the complexity of the proposed scheme to the AO algorithm, where the number of iterations of the AO algorithm is set to ${N_{i}} = 3$. It can be seen from Fig. \ref{fig42} that the complexity for performing channel estimation and reflection optimization of the proposed algorithm does not increase with the RIS size, which is due to the fact that the proposed scheme only focuses on the composite channel. By contrast, the complexity of the AO algorithm increases with the number of RIS elements $N$, which is inapplicable for scenarios with a practically large RIS. In addition, we note that although the complexity of the proposed scheme increases with the number of training slots $Q$, it is much lower than that of the AO method, which is rather applicable for practical large-scale RIS deployment.

\section{Conclusions}\label{s8}
In this paper, we proposed a scalable channel estimation and reflection optimization framework for RIS-enhanced OFDM systems. In contrast to existing solutions, the proposed scheme first generates a training set of RIS reflection coefficient vectors from the legitimate solution set. For each RIS reflection coefficient vector from the training set, the composite channel estimation and transmit power allocation inherit the same philosophy as the conventional OFDM systems operating without RIS. Thus, the RIS reflection optimization is significantly simplified upon finding out the optimal one that maximizes the achievable rate from the pre-designed training set. Furthermore, we analyzed the complexity and the achievable rate of the proposed framework, which is capable of striking flexible trade-offs between the training overhead and the achievable rate. Finally, numerical results corroborated the superiority of the proposed algorithm over the existing counterparts, especially in the presence of channel estimation errors and for rapidly time-varying channels.

\ifCLASSOPTIONcaptionsoff
 \newpage
\fi
\bibliographystyle{IEEEtran}
\bibliography{an}
\end{document}